\documentclass{PoS}
\def\be{\begin{equation}}
\def\ee{\end{equation}}
\def\dd{\displaystyle}
\def\bea{\begin{eqnarray}}
\def\eea{\end{eqnarray}}
\def\nn{\nonumber}
\def\bc{\begin{center}}
\def\ec{\end{center}}
\def\beq{\begin{equation}}
\def\eeq{\end{equation}}

\title{On Bimaximal Neutrino Mixing and GUT's}

\ShortTitle{Bimaximal Mixing and GUT's}

\author{\speaker{Guido Altarelli}\\
Dipartimento di Matematica e Fisica, Universit\`a di Roma Tre, 
\\
INFN, Sezione di Roma Tre, I-00146 Rome, Italy
\\
and
\\
CERN, Department of Physics, Theory Unit,
\\
CH-1211 Geneva 23, Switzerland\\
        E-mail: \email{guido.altarelli@cern.ch}}

\author{Pedro A. N. Machado\\
        Departamento de F\'isica Te\'orica and Instituto de F\'{\i}sica Te\'orica, IFT-UAM/CSIC,\\
Universidad Aut\'onoma de Madrid, Cantoblanco, 28049, Madrid, Spain\\
        E-mail: \email{pedro.machado@uam.es}}

\author{Davide Meloni\\
        Dipartimento di Matematica e Fisica, Universit\`a di Roma Tre, 
\\
INFN, Sezione di Roma Tre, I-00146 Rome, Italy\\
        E-mail: \email{meloni@fis.uniroma3.it}}

\abstract{We briefly discuss the present status of models of neutrino mixing. Among the existing viable options we review the virtues of Bimaximal Mixing 
(that could be implemented by an $S_4$ discrete symmetry), corrected by terms arising from the charged lepton mass diagonalization. In particular in a 
GUT formulation the property of quark lepton "weak" complementarity can be naturally realized. We discuss in some detail two new versions of particular GUT models, one based 
on $SU(5)$ and one on $SO(10)$ and the associated phenomenology. We compare these approaches based on symmetry to models based on chance, like Anarchy or $U(1)_{FN}$.
}

\FullConference{Proceedings of the Corfu Summer Institute 2014 "School and Workshops on Elementary Particle Physics and Gravity",\\
		3-21 September 2014\\
		Corfu, Greece }

\begin{document}
\section{Introduction}

A long list of models have been formulated
over the years to understand neutrino mixing. With time and the continuous
improvement of the data most of the models have been discarded by experiment. 
But the surviving models still span a wide range going from a maximum of
symmetry, as those with discrete non-abelian flavour groups 
(for  reviews, see, for example, Refs.~\cite{Altarelli:2010gt,ishikilu,grilu}), 
to the opposite extreme of Anarchy \cite{Hall:1999sn,deGouvea:2003xe,deGouvea:2012ac}.

\begin{figure}[t]
\centerline{\includegraphics[width=10 cm]{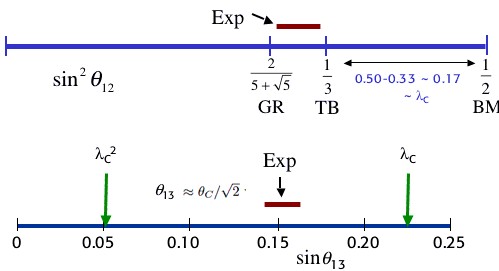}}
\caption{\it Top: the experimental value of $\sin^2{\theta_{12}}$ is compared with the predictions of exact Tri-Bimaximal (TB) or Golden Ratio (GR) or 
Bi-Maximal mixing (BM). The shift needed to bring the TB or the GR predictions to agree with the experimental value is small, 
numerically of order $\lambda_C^2$, while it is larger, of order $\lambda_C$ for the BM case, where $\lambda_C \equiv \sin{\theta_C}$ with $\theta_C$ being the 
Cabibbo angle. Bottom:  the experimental value of $\sin{\theta_{13}}$ in comparison with $\lambda_C$ or $\lambda_C^2$. \label{f1}}
\end{figure}

Among the models with a non trivial dynamical structure those based on discrete flavour groups were motivated by the fact that the data suggest some special mixing patterns as good 
first approximations like Tri-Bimaximal (TB) or Golden Ratio (GR) or Bi-Maximal (BM) mixing, for example. The corresponding mixing matrices all have 
$\sin^2{\theta_{23}}=1/2$, $\sin^2{\theta_{13}}=0$, values that are good approximations to the data (although less so since the most recent data), and differ by the value of the solar 
angle $\sin^2{\theta_{12}}$ 
(see Fig.~\ref{f1}). As the corresponding mixing matrices have the form of rotations with fixed special angles one is naturally led to discrete flavour groups.  
The relatively large measured value of $\sin{\theta_{13}}$ has disfavoured  TB and GR models because they in general predict  $\sin{\theta_{13}}$ of the same order 
as the shift of the predicted $\sin^2{\theta_{12}}$ from the observed value, which shift is small for these mixing patterns. Instead in most models of BM the measured value of 
$\theta_{13}\sim 9^\circ$ \cite{Gonzalez-Garcia:2014bfa} is natural. 

After the measurement of a relatively large value for $\theta_{13}$ there has been an intense work to interpret the new data along different approaches and ideas.  
Examples are suitable modifications of the minimal TB models \cite{lessmin,Lin:2009bw}, modified sequential dominance models  \cite{steve}, larger symmetries that  
already at LO  lead to non vanishing $\theta_{13}$ and non maximal $\theta_{23}$ \cite{altmix}, smaller symmetries that leave more freedom \cite{lesssimm}, models where 
the flavour group and a generalised CP transformation are combined in a non trivial way \cite{cpfla} (other approaches to discrete symmetry and CP violation are found in 
Refs. \cite{othercp}).

Among discrete symmetry models, now that the value of $\sin^2{\theta_{13}}$ has been measured and found to be 
not so small (\cite{Ahn:2012nd}-\cite{Abe:2013hdq}) the interest on BM mixing has been boosted \cite{Barger:1998ta,Mohapatra:1998ka}. In this case 
the neutrino mixing matrix before diagonalization of charged leptons has $\sin^2{\theta_{12}}$ and $\sin^2{\theta_{23}}$ both maximal and 
$\sin{\theta_{13}=0}$.  The BM mixing matrix is in fact:
\be
U_{BM}= \left(
\begin{array}{ccc}
\dd\frac{1}{\sqrt 2}&\dd-\frac{1}{\sqrt 2}&0\\
\dd\frac{1}{2}&\dd\frac{1}{2}&-\dd\frac{1}{\sqrt 2}\\
\dd\frac{1}{2}&\dd\frac{1}{2}&\dd\frac{1}{\sqrt 2}
\end{array}
\right)\;.
\label{BM}
\ee
The most general mass matrix corresponding to this mixing matrix is:
\be
m_{\nu BM}=\left(
\begin{array}{ccc}
x&y&y\\
y&z&x-z\\
y&x-z&z
\end{array}
\right)\;,
\label{gl2}
\ee
where $x$,$y$ and $z$ are three complex numbers.
One can think of  a suitable 
symmetry or of some other dynamical ingredient that enforces BM mixing in the neutrino sector and that the necessary, rather large, 
corrective terms to $\theta_{12}$ and $\theta_{13}$ arise from the diagonalization of the charged lepton mass matrix. As well known, 
BM corrected by charged lepton diagonalization can be obtained in models 
based on the discrete symmetry $S_4$ \cite{Altarelli:2009gn,Meloni:2011fx}. This idea is in line 
with the well-known empirical observation that $\theta_{12}+\theta_C\sim \pi/4$, where $\theta_C$ is the Cabibbo angle, a relation 
known as quark-lepton complementarity \cite{Raidal:2004iw}-\cite{Antusch:2005ca}. 
Probably the exact complementarity relation becomes more plausible if replaced by 
$\theta_{12}+\mathcal{O}(\theta_C)\sim \pi/4$ (which we call ``weak'' complementarity). In addition the measured value of 
$\theta_{13}$ is itself of order $\theta_C$:  $\theta_{13}\sim \theta_C/\sqrt{2}$.  The relation of the neutrino mixing angles 
with $\theta_C$ could well arise in Grand Unified (GUT) models  \cite{Antusch:2005ca,Patel:2010hr} so that we will focus on GUT models in this work. 

In the present note we discuss two examples of GUT models of BM, 
one based on $SU(5)$ and one on $SO(10)$. 
The $SU(5)$ model discussed here is more complete and indeed is based on a broken flavour symmetry that contains $S_4$ which imposes 
the BM structure in the neutrino sector and is then corrected by terms arising from the diagonalization of charged lepton masses.
The present $SU(5)$ model has the virtue that the quark mixing angles and the shifts from the BM 
values in the neutrino sector, all turn out to be naturally of the correct order of magnitude, expressed in terms of  powers of $\lambda_C=\sin{\theta_C}$, a' la Wolfenstein,
modulo coefficients of $\mathcal{O}(1)$. 
The $SO(10)$ model is based on Type-II see-saw and the origin of BM before diagonalization of charged leptons 
is in this case left unspecified. 
Both GUT theories discussed in the following are variants of models already appeared in the literature, 
in particular by our group \cite{Meloni:2011fx,blankenburg}. 
We discuss the phenomenology of these models in the context of the present data and the comparison with other approaches like Anarchy and $U(1)_{FN}$ models.

On the other hand, the relatively large measured value of $\theta_{13}$, close in size to the Cabibbo angle, and the indication that $\theta_{23}$ is not 
maximal both go in the direction of models based on Anarchy, i.e. the idea that perhaps no symmetry is needed in the neutrino sector, only chance. 
The appeal of Anarchy is augmented if formulated in a $SU(5) \otimes U(1)_{FN}$ context with different Froggatt-Nielsen \cite{Froggatt:1978nt} charges only for the $SU(5)$ tenplets 
(for example $10\sim(a,b,0)$, where $a > b > 0$ is the charge of the first generation, $b$ of the second, zero of the third) 
while no charge differences appear in the $\bar 5$ (e. g. $\bar 5\sim (0,0,0)$). In fact, the observed fact that the up-quark mass hierarchies are more pronounced 
than for down-quarks and charged leptons is in agreement with this assignment. Indeed the embedding of Anarchy in the $SU(5) \otimes U(1)_{FN}$ context allows to implement a 
parallel  treatment of quarks and leptons. Note that implementing Anarchy and its variants in $SO(10)$ would be difficult due to the fact that all left-handed Standard Model (SM)
fields of one generation belong to the same spinoral representation $16$.
In models with no see-saw, the $\bar 5$ charges completely fix the hierarchies (or Anarchy, if the case) in the neutrino mass matrix.  
If RH neutrinos are added, they transform as $SU(5)$ singlets and can in principle carry independent $U(1)_{FN}$ charges, which also, in the Anarchy case, 
must be all equal. With RH neutrinos the see-saw mechanism can take place and the resulting phenomenology is modified.   

The $SU(5)$ generators act vertically inside one generation, whereas the $U(1)_{FN}$ charges differ horizontally from one
generation to the other. If, for a given interaction vertex, the $U(1)_{FN}$  charges do not add to zero, the
vertex is forbidden in the symmetric limit. However, the $U(1)_{FN}$ symmetry (that one can assume to be a gauge symmetry) is spontaneously broken by
the VEVs $v_f$ of a number of {\it flavon} fields with non-vanishing charge and GUT-scale masses. Then a forbidden coupling
is rescued but is suppressed by powers of the small parameters $\lambda = v_f/M$, with $M$ a large mass, with the exponents larger for larger charge mismatch. 
Thus the charges fix the powers of $\lambda$, hence the degree of suppression of all elements of mass matrices, while arbitrary coefficients $k_{ij}$ of order 1 in each 
entry of mass matrices are left unspecified (so that the number of order 1 parameters largely exceeds the number of observable quantities). A random selection of these $k_{ij}$ 
parameters leads to distributions of resulting values for the measurable quantities. For Anarchy the mass matrices in the neutrino sector (determined by the $\bar 5$ and $1$ charges) 
are totally random, while in the presence of unequal charges different entries carry different powers of the order parameter and thus suitable hierarchies are enforced for quarks and 
charged leptons by unequal tenplet charges. 
Within this framework there are many variants of models largely based on chance: fermion charges can all be non-negative
with only negatively charged flavons, or there can be fermion charges of different signs
with either flavons of both charges or only flavons of one charge.
In Refs.\cite{AFMM:2012,melmer}, given the new experimental results, a reappraisal of Anarchy and its variants within the $SU(5)\times U(1)_{\rm FN}$ GUT framework was made. 
Based on the most recent data it is argued  that the Anarchy ansatz is probably oversimplified and, in any case, not compelling. 
In fact, suitable differences of $U(1)_{FN}$ charges, if also introduced within pentaplets and singlets, lead, with the same number of random parameters as for Anarchy, 
to distributions that are in better agreement with the data.

%%%%%%%%%%%%%%%%%%%%%%%%%%%%%%%%%%%%%%%%%%%%%%%%%%%%%%%%%%%%%%%%%
\section{A SUSY $SU(5)$ model with $S_4$ discrete symmetry}
\label{sec:$SU(5)$}

This model is a variant of those of Refs.\cite{Altarelli:2009gn,Meloni:2011fx} to which we refer the reader for a detailed discussion and 
the technical details \cite{5DSU5}-\cite{Altarelli:2008bg}. 
Here we concentrate on the differences, the motivations and the resulting phenomenology. This is a SUSY $SU(5)$ model in 4+1 
dimensions with a flavour symmetry $S_4 \otimes Z_3 \otimes U(1)_R \otimes U(1)_{FN}$, where $U(1)_R$ implements the 
R-symmetry while $U(1)_{FN}$ is a Froggatt-Nielsen (FN) symmetry \cite{Froggatt:1978nt} that induces the hierarchies of fermion masses 
and mixings. 
The particle assignments are displayed in Tab.\ref{tab:Multiplet1}. 

\begin{table}[h!]
\begin{center}
\begin{tabular}{|c||c|c|c|c|c|c|c|c|c|c|c|c|c|c|c|c||}
\hline
{\tt Field} & $F$ & $T_1$ & $T_2$ & $T_3$ & $H_5$ & $H_{\overline 5}$ &   $\varphi_\nu$ & $\xi_\nu$ &  $\varphi_\ell$& $\chi_\ell$ &  
$\theta$ &  $\theta^\prime$ & 
$\varphi^0_\nu$& $\xi^0_\nu$ &  $\psi^0_\ell$& $\chi^0_\ell$ 
\\
\hline\hline
SU(5) & $\bar{5}$  & 10 & 10 & 10 & 5 & ${\overline {5}}$ &  1 & 1 & 1 & 1 & 1 & 1 & 1 & 1 & 1 & 1  \\
\hline
$S_4$  &  $3_1$ & 1  & 1 & 1 & 1 & 1 &   $3_1$ & 1 & $3_1$ & $3_2$ &  1 & 1 & $3_1$ & 1 & 2 & $3_2$     \\
\hline
$Z_3$ & $\omega$ & $\omega$ &1  &  $\omega^2$ & $\omega^2$ &$\omega^2$ &  1 & 1 & $\omega$ &$\omega$ &1 &  $\omega$ & 1 & 1 & 
$\omega$ & $\omega$  \\
\hline
$U(1)_R$ & 1 & 1 & 1 & 1 & 0 & 0&  0 & 0 & 0 & 0& 0 & 0 & 2 & 2 & 2 & 2 \\
\hline
$U(1)_{FN}$ & 0 & 2 & 1 & 0 & 0 &  0 & 0 & 0 & 0& 0 & -1 & -1 & 0 & 0 & 0 & 0 \\
\hline 
  & {\tt br} &  {\tt bu} & {\tt bu} & {\tt br} &  {\tt bu} &  {\tt bu} &   {\tt br} &  {\tt br} &  {\tt br} & 
 {\tt br} &  {\tt br} &  {\tt br} &  {\tt br} &  {\tt br} &  {\tt br} &  {\tt br}\\
\hline 
\end{tabular}
\caption{\label{tab:Multiplet1}\it Matter assignment of the model. The symbol ${\tt br}({\tt bu})$ indicates that the corresponding 
fields live
on the brane (bulk).}
\end{center}
\end{table}

The formulation in 4+1 dimensions,  with coordinate $y$ in the fifth dimension compactified on a circle, allows a more efficient 
realization of GUT with less Higgs states, no doublet-triplet splitting problem and a better control of proton decay. In the present
model it also introduces some extra hierarchy for some of the couplings.  In fact, as indicated in the table, some of the particles 
are in the bulk (the first two generation tenplets $T_1$ and $T_2$ and the Higgs $H_5$ and $H_{\bar 5}$) while all the other ones are 
on the brane at $y=0$. Actually, to obtain the correct zero mode spectrum, one must introduce two copies, $T_{1,2}$ and $T^\prime_{1,2}$ 
with opposite orbifolding parity.  The zero modes of $T_{1,2}$ are the quark doublets $Q_{1,2}$, while those of $T^\prime_{1,2}$ are 
$U^c_{1,2}$ and $E^c_{1,2}$. For economy of space only $T_{1,2}$ appear in table 1. At leading order (LO) the $S_4$ symmetry is broken 
down to suitable different subgroups in the charged lepton sector and in the neutrino sector by the VEV's of the flavons 
$\varphi_\nu$, $\xi_\nu$,  $\varphi_\ell$ and $\chi_\ell$. The necessary proper alignment of the VEV's is implemented, in a natural 
way, by the driving fields $\varphi^0_\nu$,  $\xi^0_\nu$,  $\psi^0_\ell$, $\chi^0_\ell$. The VEVs of the $\theta$ and $\theta^\prime$ 
fields break the FN symmetry.  As a result, at LO the charged lepton masses are diagonal and exact BM is realized for neutrinos.  
Corrections to diagonal charged leptons and to exact BM are induced by vertices of higher dimension in the Lagrangian, suppressed by 
powers of a large scale $\Lambda$. While broken $S_4$ is at the origin of BM, $U(1)_{FN}$, together with some higher dimensional effects, fixes the hierarchies of quark and charged 
lepton masses. 

With respect to Ref.\cite{Meloni:2011fx} here we have modified the FN charges of the tenplets from (3,1,0) to (2,1,0). 
With this choice we aim at 
optimizing the neutrino mixing angles and to implement the "weak complementarity" relation at the expenses of a less accurate 
description of the first generation quark and charged lepton masses. We adopt the definitions:
\bea
\frac{v_{\varphi_\ell}}{\Lambda} \sim \frac{v_{\chi}}{\Lambda} \sim \epsilon^\prime\;; 
\qquad \frac{v_{\varphi_\nu}}{\Lambda} \sim \frac{v_{\xi}}{\Lambda} \sim \epsilon\;,
\eea
\bea
\frac{\langle \theta \rangle}{\Lambda} = t \qquad \frac{\langle \theta^\prime \rangle}{\Lambda} = t^\prime\,.
\eea 
and
\be
s\equiv\dd\frac{1}{\sqrt{\pi R \Lambda}}<1~~~.
\label{vsup}
\ee  
where $s$ is the volume suppression factor. 

It turns out that the simplest choice of setting all these parameters to be of $\mathcal{O}(\lambda \sim \lambda_C)$: 
\be
s=\epsilon=\epsilon^\prime=t=t^\prime=\lambda_C, 
\label{equal}
\ee
leads to a good description of masses and mixings, as described below. 
Indeed by proceeding in exact analogy with Ref.\cite{Meloni:2011fx}, we have the following results.

\subsection{Down quarks and charged lepton mass matrices}
\label{downq}

For the down quark mass matrix, by only keeping the leading terms for each entry,  one finds:

\be
m_d=\left(
\begin{array}{ccc}
a_{11}\lambda^5&a_{12} \lambda^4& a_{13} \lambda^4\\
a_{21} \lambda^4&-c \lambda^3& c\lambda^3\\
a_{31}\lambda^2& ......&a_{33} \lambda
\end{array}
\right) \lambda v_d^0~~~.
\label{mdlam}
\ee
Here all matrix elements are multiplied by generic coefficients $a_{ij}$ of $\mathcal{O}(1)$ with the exception of the (22) and (23) 
entries 
where the coefficients given by $-c$ and $c$ are equal and opposite. In the (32) entry the dots indicate that the lowest order potentially 
non vanishing is actually zero and the entry will get a contribution from still higher orders. For example, the entry (11) gets 
contributions from terms $\mathcal{O}(st \epsilon^\prime t^\prime \epsilon \sim \lambda^5)$. Thus the hierarchies arise from a 
combination of extra dimension factors, $S_4$ and $U(1)_{FN}$ breaking. For the mass eigenvalues we have:
\be
m_b \sim v_d^0 \lambda^2,~~~~~~~m_s \sim v_d^0 \lambda^4,~~~~~~~m_d \sim v_d^0 \lambda^6.
\ee

For the charged lepton masses we have to take into account the  introduction of the copies 
$T^\prime_{1,2}$ of the first two tenplet fields, whose zero modes are different from 
those of $T_{1,2}$ and couple with the charged leptons only. Therefore, all the operators 
of the form $F T_3 H_{\overline 5}$ have exactly the same order 1 coefficients whereas all others containing 
$T^\prime_{1,2}$ have different coefficients. This translates in the following mass matrix:
\be
m_e=\left(
\begin{array}{ccc}
a_{11}^\prime\lambda^5&a_{21}^\prime \lambda^4& a_{31}\lambda^2\\
a_{12}^\prime \lambda^4&-c^\prime \lambda^3& ...... \\
a_{13}\lambda^4&c^\prime \lambda^3&a_{33} \lambda
\end{array}
\right) \lambda v_d^0~~~,
\label{melam}
\ee
and the  charged lepton masses are:
\be
m_\tau=m_b \sim v_d^0 \lambda^2,~~~~~~~m_\mu \sim v_d^0 \lambda^4,~~~~~~~m_e \sim v_d^0 \lambda^6.
\ee
Note that the $b-\tau$ universality is realized: here and in the following we obviously refer to masses at the GUT scale as for example 
given in Ref.\cite{Joshipura:2011nn}. Note that the predicted ratio $m_e/m_\mu \sim \lambda^2$ is not perfect, being too large.

The unitary left-handed rotation $U_d$ is obtained diagonalizing $m_d\,m_d^\dagger$ whereas the right-handed one $U_\ell$
is the charged lepton rotation. Taking only the largest contribution for each matrix elements, for $U_\ell$, which enters in the neutrino 
mixing matrix, $U_{PNMS}= U_\ell^\dagger U_\nu$ we have:

\bea
\label{ul}
U_\ell \sim  
\left(
\begin{array}{ccc}
1  &u_{12}\lambda&u_{13}\lambda  \\
-u_{12}^* \lambda&  1 & 0 \\
-u_{13}^* \lambda   &  
-u_{12}^* u_{13}^*\lambda^2     & 1 \\
                     \end{array}
                   \right)\,,
\eea 
so that $\theta_{23}^\ell = 0$ in this approximation.

\subsection{Up quarks  mass matrix}
\label{upq}

Similarly the symmetric up quark mass matrix is given by:
\be
m_u=\left(
\begin{array}{ccc}
b_{11}\lambda^6&b_{12} \lambda^5& b_{13} \lambda^3\\
b_{12} \lambda^5&b_{22}\lambda^4&b_{23}\lambda^4\\
b_{13}\lambda^3&b_{23}\lambda^4&b_{33}
\end{array}
\right) \lambda v_u^0~~~,
\label{mulam}
\ee
and the masses are:
\be
m_t \sim v_u^0 \lambda,~~~~~~~m_c \sim v_u^0 \lambda^5,~~~~~~~m_u \sim v_u^0 \lambda^7.
\ee
The ratio $m_b/m_t \sim \lambda v_d^0/v_u^0 \sim \lambda / \tan\beta$ implies that $\tan\beta \sim 1/\lambda$. In most of the cases the 
mass ratios are correctly reproduced, but, as already announced,  there are some cases that need some moderate fine tuning. For example,
we have $m_u/m_c \sim \lambda^2$ which is too large, $m_u/m_d \sim m_c/m_s $ where, in reality the two sides differ by a factor of 
about 25. As we already mentioned we have chosen as a priority to fit the mixings rather than the masses. 
One can improve the agreement by somewhat relaxing the strict equalities in eq. (\ref{equal}).
We now turn to 
describe the model predictions for the mixings first in the quark sector and then, after discussing the neutrino mass matrix, in the 
leptonic sector.

\subsection{The CKM matrix}
\label{vckm}

The CKM matrix is given by $V_{CKM}=U_u^\dagger U_d$. The leading order expressions for $U_u$ and $U_d$ obtained from the up and down 
quark mass matrices in eqs.(\ref{mdlam}) and (\ref{mulam}) are of the form:
\bea
\label{uu}
U_u \sim  
\left(
\begin{array}{ccc}
1  &c_{12}\lambda&u_{13}\lambda^3  \\
-c_{12}^* \lambda&  1 & 0 \\
-c_{13}^* \lambda^3   &  
0& 1 \\
                     \end{array}
                   \right)\,
\eea 
and 
\bea
\label{ud}
U_d \sim  
\left(
\begin{array}{ccc}
1  &d_{12}\lambda&d_{13}\lambda^3  \\
-d_{12}^* \lambda&  1 & d_{23}\lambda^2 \\
(d_{12}^*d_{23}^* -d_{13}^*)\lambda^3   &  
-d_{23}\lambda^2 & 1 \\
                     \end{array}
                   \right)\,.
\eea 
From these expressions we obtain the leading order form of the $V_{CKM}$ matrix with a pattern of the Wolfenstein type:
\bea
\label{ckm}
V_{CKM} \sim  
\left(
\begin{array}{ccc}
1  &v_{12}\lambda&v_{13}\lambda^3  \\
-v_{12}^* \lambda&  1 & v_{23}\lambda^2 \\
(v_{12}^*v_{23}^* -v_{13}^*)\lambda^3   &  
-v_{23}\lambda^2 & 1 \\
                     \end{array}
                   \right)\,.
\eea 
The $v_{ij}$ coefficients are related to the $c_{ij}$ and $d_{ij}$ coefficients by:
\be
v_{12}=d_{12}-c_{12}^*;~~~~~v_{13}=(d_{13}-c_{12}^*d_{23}-c_{13}^*);~~~~~v_{23}=d_{23}.
\ee
We see that for $\lambda = \lambda_C$ the correct order of magnitude is derived for each  $V_{CKM}$ matrix element modulo coefficients 
generically of order 1.
\subsection{The neutrino masses and mixings}
\label{neutrinos}

The neutrino sector of the model is unchanged with respect to Ref.\cite{Meloni:2011fx}. 
We therefore limit ourselves here to recall some important points. First note that in Tab.\ref{tab:Multiplet1} 
there are no right-handed neutrinos. 
So the table refers to a model where the neutrino mass matrix is generated by the effective dimension 5 Weinberg operator. 
But a see-saw version is easily obtained by adding 3 right-handed neutrinos  transforming under SU(5)$\times S_4\times Z_3$ 
as $(1,3_1,1)$ and with charges $U(1)_R=+1$ and $U(1)_{FN}=0$. The relevant phenomenology is quite similar, in particular the  results on neutrino mixing.  
However, the neutrino mass spectrum turns out to be of moderate normal hierarchy type, with a LO lightest neutrino mass $|m_1|$ larger than about 0.01 eV and, 
consequently, values of $m_{ee} \geq 3 - 4~10^{-3}$ eV. The deviation from a pure BM neutrino mass matrix is responsible for a softening of the lower bound on $|m_1|$ and 
for an enlargement of the allowed values of the $0\nu\beta\beta$ rate. The experimental value of 
$\sqrt{r} =\sqrt{ \frac{\Delta m^2_{sol}}{\Delta m^2_{atm}}} \sim 1/6$ needs some fine tuning 
because in the model it should be of $\mathcal{O}(1)$. This fine tuning appears in most discrete symmetry models because neutrinos 
must be in triplet representations of the discrete group in order to obtain BM or TB mixing etc. and then the mass eigenvalues are all of 
the same order of magnitude, barring cancellations.  

The neutrino mixing matrix is obtained as $U_\nu = U_\ell^\dagger U_{BM}$ where  $U_\ell$ is given in eq. (\ref{ul}) and $U_{BM}$ is 
the unitary matrix of BM. The results for the mixing angles are easily derived:
\bea
\sin{\theta_{13}} &=&  \frac{1}{\sqrt{2}} |u_{12}-u_{13}|\lambda \equiv |\Delta| \nn \\
\sin^2{\theta_{12}} &=& \frac{1}{2}- \frac{1}{\sqrt{2}}~Re(u_{12}+u_{13})\lambda \equiv \frac{1}{2}- Re \Sigma \label{final2}  \\
\sin^2{\theta_{23}} &=&\frac{1}{2}+ \mathcal{O}(\lambda^2) \nn
\eea
The CP phase $\delta_{CP}$ is
\be
\delta_{CP}= \pi + arg(u_{12}-u_{13}) \equiv \pi + arg(\Delta).
\ee
where the complex numbers $\Delta$ and $\Sigma$ are defined as $\Delta= \frac{1}{\sqrt{2}} |u_{12}-u_{13}|\lambda$ and 
$\Sigma=\frac{1}{\sqrt{2}}~(u_{12}+u_{13})\lambda$.
The results are graphically reported in Fig. 2 and compared with the experimental values of $\sin{\theta_{13}}$ and $\sin^2{\theta_{12}}$.
We see that, with $\lambda \sim \lambda_C$, the model realizes the "weak" complementarity relation and the experimental fact that $\sin{\theta_{13}}$ is of the same order 
than the shift of 
$\sin^2{\theta_{12}}$ from the BM value of 1/2, both of order $\lambda_C$. 
It is interesting to observe that corrections to the BM pattern arising from next-to-leading order effects 
in the Yukawa couplings (higher order operators and shifts from LO flavon vevs) only affect 
$\sin^2{\theta_{23}}$ at the same $\mathcal{O}(\lambda^2)$ as in eq.(\ref{final2}). We also see that, in general, the CP phase $\delta_{CP}$ is not predicted, 
as the  data only fix the absolute value of $\Delta$ and not its phase. If one could neglect $u_{13}$ then $\Delta$ and $\Sigma$ would coincide. 
A very marginal agreement with the data would then demand that both be aligned along the positive real axis and, in this case $\delta_{CP}= \pi$.

% A see-saw version of the model is easily obtained by adding 3 right-handed neutrinos  transforming under SU(5)$\times S_4\times Z_3$ 
% as $(1,3_1,1)$ and with charges $U(1)_R=+1$ and $U(1)_{FN}=0$. The relevant phenomenology is quite similar to the no see-saw case, 
% in particular the results on neutrino mixing given in eq.(\ref{final2}) are confirmed.  
% However, the neutrino mass spectrum turns out to be of moderate normal hierarchy type \cite{Altarelli:2009gn}, 
% with a  LO  lightest neutrino mass $|m_1|$ larger than about 0.01 eV and, consequently, 
% values of $|m_{ee}| \gtrsim 3-4 \cdot 10^{-3}$ eV. Deviation from a pure BM neutrino mass matrix 
% is responsible for a softening of the lower bound on $|m_1|$ and for an enlargement of the allowed values  
% of the $0\nu\beta \beta$ rate.

\begin{figure}[h!]
\bc
\includegraphics[scale=.4]{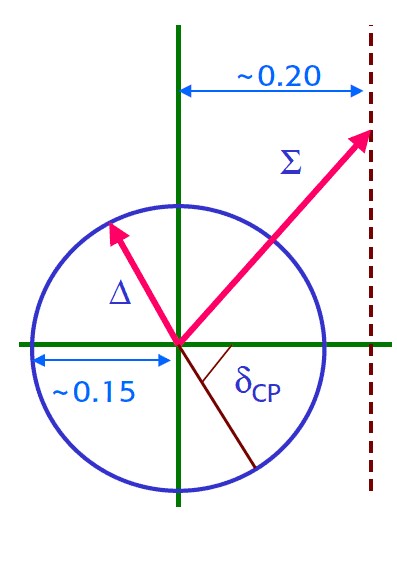}
\caption{\label{fig1}\it The complex numbers $\Delta= \frac{1}{\sqrt{2}} |u_{12}-u_{13}|\lambda$ and 
$\Sigma=\frac{1}{\sqrt{2}}~(u_{12}+u_{13})\lambda$ determine  $\sin{\theta_{13}}=|\Delta|$ and 
$\sin^2{\theta_{12}}= \frac{1}{2}- Re \Sigma $. The CP phase is given by $\delta_{CP}= \pi + arg(\Delta)$. The numbers shown approximately indicate the experimental central values.}
\ec
\end{figure}

\section{Comparison with $SU(5)\bigotimes U(1)$ models }
\label{sec:FN}

It is interesting to compare the previous model, which is rather complicated involving SUSY $SU(5)$, 
a non abelian $S_4$ symmetry and extra dimensions, with a much simpler class of models based on SUSY $SU(5) \otimes U(1)_{FN}$
\cite{Froggatt:1978nt}. 
As we have explicitly discussed a non see-saw version of the $S_4$ 
model we will compare it with a non see-saw version of the $U(1)_{FN}$ models. 
In the following,  only $U(1)_{FN}$ models with normal hierarchy are considered because, as shown in Ref.\cite{Altarelli:2002sg},   
$U(1)$ models with inverse hierarchy (IH) tend to favour a solar angle close to maximal.

In general we can label the $U(1)$ charges as follows:
\be 
10 \sim (t_1, t_2, 0) ~~~~~~~~~~~~ \bar 5 \sim (f_1, f_2, 0)\\
\ee
where 1, 2 refer to the first and second families. The Higgs field charges are taken as vanishing. 
There is also a flavon field $\theta$ with charge -1 whose VEV $\langle \theta \rangle$ breaks $U(1)_{FN}$. 
A set of charge values that lead to a good agreement with the observed masses and mixings are (this is the so-called 
{\it $A_{\mu\tau}$} case in \cite{Altarelli:2012ia}):
\be 
10 \sim (3, 2, 0) ~~~~~~~~~~~~ \bar 5 \sim (1, 0, 0)\,.\\
\ee
The following mass matrices are obtained. For the up-type quarks:
\be
m_u=\left(
\begin{array}{ccc}
 \lambda^6&\lambda^5&\lambda^3\\
 \lambda^5&\lambda^4&\lambda^2\\
\lambda^3&\lambda^2&1
\end{array}
\right) v_u^0~~~.
\label{mulamprime}
\ee
Here $\lambda = \langle \theta\rangle/\Lambda$ with $\Lambda$ the large scale that suppresses the non rinormalizable interactions involving 
the field $\theta$ and all entries are multiplied by coefficients that are complex numbers with absolute values of order 1.

The down-type quarks and charged lepton mass matrices are one the transposed of the other and are given by:
\be
m_d=m_e^T=\left(
\begin{array}{ccc}
\lambda^4& \lambda^3&  \lambda^3\\
 \lambda^3&\lambda^2& \lambda^2\\
 \lambda&1 &1
\end{array}
\right) v_d^0~~~.
\label{mdlamprime}
\ee

Finally, the neutrino mass matrix transforming as $\bar 5 \otimes \bar5$ from the dimension 5 Weinberg operator is given by:

\be
m_\nu=\left(
\begin{array}{ccc}
 \lambda^2&\lambda&\lambda\\
 \lambda&1&1\\
\lambda&1&1
\end{array}
\right) \frac{(v_u^{0})^{2}}{\Lambda}~~~.
\label{nuprime}
\ee

We can now compare the two models starting from quarks and charged leptons. We observe that the  structure of the mass matrices is 
very similar although not identical.  
In the $S_4$ model the suppression factors from the geometry in the extra dimension combine with those from the $U(1)_{FN}$ charges 
(which are different from those in the abelian model) to produce a similar pattern in the two cases. The mass matrices are not 
precisely the same in the two models but the predictions for the orders of magnitude of the mass ratios, expressed in terms of powers 
of $\lambda$ are identical. For both models we have in fact:
\bea 
\frac{m_u}{m_c}\sim \frac{m_d}{m_s}\sim \frac{m_s}{m_b}&\sim &\frac{m_e}{m_\mu}\sim \frac{m_\mu}{m_\tau}\sim \lambda^2\\
\frac{m_c}{m_t}&\sim & \lambda^4
\eea
Most of these orders of magnitude are correct if $\lambda \sim \lambda_C$ but some, involving the first generation masses, are not, 
like $m_u/m_c$, $m_d/m_s$ and $m_e/m_\mu$. Although the predictions for the mass ratios are the same,
the $S_4$ model is superior, because its extra dimensional formulation solves the doublet triplet splitting problem and introduces 
corrections to the relation $m_d = m_e^T$ of the Georgi-Jarlskog type. Also in the $S_4$ model 
$m_t/m_b \sim v^0_u/(\lambda v^0_d)\sim \tan \beta/\lambda$ while the factor $1/\lambda$ is absent in the $U(1)_{FN}$ model, 
so that only a moderate value of $\tan\beta$ is needed in the $S_4$ model. The CKM quark mixing angles are also of the same order 
of magnitude in the two models and match the Wolfenstein pattern: $\theta_{12}^q \sim \lambda$, $\theta_{13}^q \sim \lambda^3$ and 
$\theta_{23}^q \sim \lambda^2$.

In conclusion, in the charged fermion sector the two models are rather comparable, with some advantages for the $S_4$ model. But where the 
latter is definitely superior is in the neutrino sector. As a result of the $S_4$ construction, the neutrino mixing pattern is 
dictated by BM corrected by terms from the diagonalization of charged leptons as detailed in eq.(\ref{final2}). 
The weak form of complementarity is 
realized as the shift of  $\theta_{12}^\nu$ is of the order of $\lambda \sim \lambda_C$ and moreover  also 
$\theta_{13}^\nu \sim \lambda_C$ while $\theta_{23}^\nu$ deviates from the maximal value by terms of order $\lambda^2$. 
As already mentioned the observed value of $\sqrt{r} =\sqrt{ \frac{\Delta m^2_{sol}}{\Delta m^2_{atm}}} \sim 1/6$ needs some fine tuning because in 
the model it should be of $\mathcal{O}(1)$. 
In the $U(1)_{FN}$ model the neutrino matrix is given in eq.(\ref{nuprime}). The diagonalization of charged 
leptons does not alter this pattern. For generic coefficients of $\mathcal{O}(1)$ for each matrix entry, we would get that 
$\theta_{13}^\nu \sim \lambda$, $\theta_{23}^\nu \sim \mathcal{O}(1)$ which are good but also $r \sim \mathcal{O}(1)$ and 
$\theta_{12}^\nu \sim \lambda$ which are bad. However, 
if, by accident, the 22 matrix element of $m_\nu$ is of order $\lambda$, then  
$\sqrt{r} \sim \lambda$ and $\theta_{12}^\nu \sim \mathcal{O}(1)$: with a single fine tuning one fixes both problems. 
In any case, even accepting some amount of fine tuning, clearly there is no realization of weak complementarity. 
Finally in both models there are far more 
parameters than observables. This redundancy is less pronounced for the $S_4$ model but is still large.

\section{Bimaximal mixing in a $SO(10)$ GUT model}
\label{sec:bmso10}

A challenging problem is that of formulating a natural model of Grand Unification based on $SO(10)$, leading not only to a good description of quark masses 
and mixing but also, in addition, of charged lepton masses and neutrino mixing. In $SO(10)$ the main added difficulty with respect to $SU(5)$ is clearly that all fermions 
in one generation belong to a single 16-dimensional representation, so that one cannot separately play with the properties of the $SU(5)$-singlet right-handed neutrinos in 
order to explain the striking difference between quark and neutrino mixing. A promising strategy in order to separate charged fermions and neutrinos in $SO(10)$ is to assume 
the dominance of type-II see-saw \cite{tII} (with respect to type-I see-saw \cite{tI}) for the light neutrino mass matrix. If type-II seesaw is responsible for neutrino
masses, then the neutrino mass matrix (proportional to) $f$ (see eqs.(\ref{eq1},\ref{vr},\ref{eq4})) is separated from the dominant contributions to the charged fermion 
masses and can therefore show a completely different pattern. This is to be compared with the case of type-I see-saw where the neutrino mass matrix depends on the neutrino 
Dirac and Majorana matrices and, in $SO(10)$, the relation with the charged fermion mass matrices is tighter.

In Ref.\cite{blankenburg}, when the data suggested approximate TB mixing and a small value of $\sin^2{\theta_{13}}$, an $SO(10)$ model has been studied based on type-II 
see-saw dominance. 
A detailed discussion of the general structure of this class of models can be found in the above article, together with a comparison with other approaches to $SO(10)$ GUT's. 
Here, given the relatively large value of $\sin^2{\theta_{13}}$  that has been recently measured, we reconsider this type of GUT model in the case of approximate BM mixing 
corrected by the charged lepton diagonalization. 

In renormalizable $SO(10)$ models (a non necessary assumption only taken here for simplicity) the Higgs fields that
contribute to fermion masses are in {\bf 10} (denoted by $H$), ${\bf \overline{126}}$
($\overline{\Delta}$) and {\bf 120} ($\Sigma$). 
The Yukawa
superpotential of this model is then given by:
\begin{eqnarray}
W_Y~=~h\, \psi\psi H + f\, \psi\psi\bar{\Delta}+h'\,\psi\psi \Sigma\,,
\label{eq1}
\end{eqnarray}
where the symbol $\psi$ stands for the {\bf 16} dimensional
representation of SO(10) that includes all the fermion fields in one generation.
The coupling matrices $h$ and $f$ are symmetric,
while $h^\prime$ is anti-symmetric.
The representations $H$ and $\Delta$ have two SM doublets in each of them whereas $\Sigma$ has four such doublets.
At the GUT scale $M_{GUT}$,
once the GUT and the $B-L$ symmetry are broken, one linear
combination of the up-type and one of down-type doublets
remain almost massless whereas the remaining combinations acquire GUT
scale masses.
The electroweak symmetry is broken after the light Minimal Supersymmetric Standard Model (MSSM) doublets
(to be called $H_{u,d}$) acquire vacuum expectation values (vevs) and
they then generate the
fermion masses. The resulting mass formulae for  different
fermion masses are given by (see, for example, \cite{mimura}):
\begin{eqnarray}
Y_u &=& h + r_2 f +r_3 h^\prime, \label{eq2} \\\nonumber
Y_d &=& r_1 (h+ f + h^\prime)\,, \\\nonumber
Y_e &=& r_1 (h-3f + c_e h^\prime)\,, \\\nonumber
Y_{\nu^D} &=& h-3 r_2 f + c_\nu h^\prime,
\end{eqnarray}
where $Y_a$ are mass matrices divided by the electro-weak vev's
$v_{u,d}$ and $r_a$ ($a=1,2,3$ and $c_b$ ($b=e,\nu$) are the mixing parameters which
relate the $H_{u,d}$ to the doublets in the various GUT
multiplets.

In generic $SO(10)$ models of this type, the neutrino
mass formula has a type-II and a type-I contribution:
\begin{eqnarray}
{\cal M}_\nu~=~fv_L-M_D\frac{1}{fv_R}M^T_D\,,
\label{vr}
\end{eqnarray}
where $v_L$ is the vev of the $B-L=2$ triplet in the ${\bf\overline{126}}$ Higgs
field. Note that in general, the two
contributions to neutrino mass depend on two different parameters, $v_L$ and $v_R$,
and it is possible to have a symmetry breaking pattern in
$SO(10)$ such that the first contribution (the type-II term)
dominates over the type-I term. The possible realisation of this dominance and its consistency with coupling unification has been studied 
in the literature 
\cite{goh, aulak1, aulak2, bajc1} and found tricky but not impossible \cite{melfo}. The neutrino mass formula then
becomes
\begin{eqnarray}
{\cal M}_\nu~\sim~fv_L\;.
\label{eq4}
\end{eqnarray}
Note that $f$ is the same coupling matrix that appears in the
charged fermion masses in eq. (\ref{eq2}),
up to factors from the Higgs mixings and the Clebsch-Gordan coefficients. Also note that the neutrino Dirac mass, proportional to $Y_{\nu^D}$ in eq. (\ref{eq2}), 
only enters in the neglected type-I see-saw terms and does not play a role in the following analysis.
The equations (\ref{eq2}) and (\ref{eq4}) are
the key relations in this approach. 

The {\bf 10} Yukawa couplings
contributing to up, down and charged lepton masses in most models have a large 33 term, corresponding to the large third generation masses, while all other entries are 
smaller and lead by themselves to zero CKM mixing (because the {\bf 10} contributes equally to up and down mixing). Quark mixings arise from small corrections due to
${\bf \overline{126}}$, the same Higgs representation that determines $f$ which in models with type-II see-saw is dominant in the neutrino sector, and to {\bf 120}.  
Thus, in this approach, in the absence of  {\bf 120}, there is a strict relation between quark masses and mixings and the neutrino mass matrix. The presence of {\bf 120} 
dilutes this connection which however still remains important. In particular the deviations from BM mixing induced by the diagonalization of the charged lepton mass matrix, 
are typically of the same order as the largest quark mixing angle i.e the Cabibbo angle.  An interesting question is to see to which extent the data are compatible with the 
constraints implied by this interconnected structure. 

For generic eigenvalues $m_i$, the most general matrix that is diagonalized by the BM unitary transformation is given by:
\beq
f=U_{BM}^* {\rm diag}(m_1,m_2,m_3) U_{BM}^\dagger~~~.
\label{numass}
\eeq
where $U_{BM}$ is the BM mixing matrix given in eq.(\ref{BM}). In this convention $U_{BM}$ is a real orthogonal matrix and all phases can be included in the eigenvalues $m_i$.  
Then the matrix $f$ is symmetric with complex entries and,
from eq. (\ref{numass}), one obtains (see eq.(\ref{gl2})):
\be
f=\left(
\begin{array}{ccc}
f_2&f_1&f_1\\
f_1&f_0&f_2-f_0\\
f_1&f_2-f_0&f_0\
\end{array}
\right)\;,
\label{effe}
\ee
with: $m_1=f_2+\sqrt{2}f_1$, $m_2=f_2-\sqrt{2}f_1$ and $m_3=2f_0-f_2$.

An important observation is that, for a generic neutrino mass matrix $f'$, we can always go to a basis where $f'$ is diagonalized by the BM unitary transformation 
in eq. (\ref{BM}) and is of the form in eq. (\ref{effe}), in the same way as discussed in Ref. \cite{blankenburg} for TB mixing. In fact, if we start from a complex symmetric 
matrix $f'$ not of that form, 
it is sufficient to diagonalize it by a unitary transformation $U$: $f'_{diag}=U^Tf'U$ and then take the matrix 
\begin{equation}
f=U_{BM}^*f'_{diag} U_{BM}^\dagger=U_{BM}^* U^T f' U U_{BM}^\dagger\,.
\label{efefpr}
\end{equation} 
As a result the matrices $f$ and $f'$ are related by a change of the charged lepton basis induced by the unitary matrix $O= U U_{BM}^\dagger$ (in $SO(10)$ the matrix $O$ rotates the 
 whole fermion representations $\bf 16_i$).  Since BM mixing is not a very good approximation to the data, in this basis substantial deviations from BM mixing must be generated by the 
diagonalization of charged leptons, with terms expected to be of  $\mathcal{O}(\lambda_C)$.  At the same time also the quark mixings must be reproduced in agreement with the data. 
As the matrix elements of $f$ enter both in the neutrino mass formula and in the corrections to the fermion mass matrices, this fact poses a non trivial problem of consistency, especially in view of the small values of the first generation masses.

As one could decide to work in a basis where the matrix $f$ is diagonalised by the TB matrix or by BM matrix (or in another suitable basis), this means that, for the measured set of data, the result of a best fit performed in one basis should lead to the same $\chi^2$ than the fit in other basis, because the only difference is that the set of parameters used in one fit are functions of the parameters of the other fit. So the $\chi^2$ cannot decide whether TB or BM is a better starting point. However, since the first generation masses are very small some parameters must be precisely fine tuned in order to reproduce the small values of the masses. It is possible that one needs more fine tuning in one case than in the other. For a quantitative measure, in a given fit,  of the amount of fine-tuning needed a parameter $d_{FT}$ was introduced in Ref. \cite{blankenburg}. 
This adimensional quantity is obtained as the sum of the absolute values of the ratios between each parameter $p_i$ and its "error", defined, for this purpose, as the shift from the best fit value that changes the $\chi^2$ by one unit, with all other parameters fixed at their best fit values (this is not the error given by the fitting procedure because in that case all the parameters are varied at the same time and the correlations are taken into account):
\begin{equation}
\label{fine-tuning}
d_{FT} = \sum\mid \frac{par_i}{err_i} \mid 
\end{equation}
It is clear that $d_{FT}$ gives a rough idea of the amount of fine-tuning involved in the fit because if some $|err_i/par_i|$ are very small it means that it takes a minimal variation of the corresponding parameter to make a large difference on the $\chi^2$. 

%Another possible measure of fine tuning is related to an average convexity of the $\chi^2$ with respect to the parameters:
%\begin{equation}
%\label{fine-tuning2}
%\delta_{FT} =\sum_{ij}  \mid \frac{\chi^2}{\partial par_i \partial par_j} \mid 
%\end{equation}

We report here on a comparative study of starting from $f$ in the TB or in the BM basis. For the TB case the important difference with the detailed, complete discussion in Ref.\cite{blankenburg} is that here we used updated experimental values for the neutrino mixing angles, 
in particular for $\sin^2\theta_{13}\sim 0.022\pm0.001$, as most precisely measured by the  Daya Bay experiment \cite{An:2013zwz}. The result of a best fit performed in one basis should lead to the same $\chi^2$ than the fit in another basis, because the only difference is that the set of parameters 
used in one fit are functions of the parameters of the other fit. So, as we have already stressed, the $\chi^2$ cannot decide whether TB or BM is better. We have checked that the $\chi^2$ is equal within uncertainties in the two cases, and this is true even for values of $\sin\theta_{13}$ somewhat different than the measured value, as can be seen in Fig. 3. However, since the first generation masses are very small 
some parameters must be precisely fine tuned in order to reproduce the small values of the masses. 
% It is possible that one needs more fine tuning in one case than in the other. 
It turns out that, for the physical value of $\sin^2\theta_{13}$, $d_{FT}$ is smaller in the TB case. A study of the fine tuning parameter when the fit is repeated with the same data except for $\sin^2\theta_{13}$, which is moved from small to large, shows that the fine tuning increases (decreases) with $\sin\theta_{13}$ for TB (BM), as shown in Fig. 4.

\begin{figure}[h!]
\bc
\includegraphics[scale=.5]{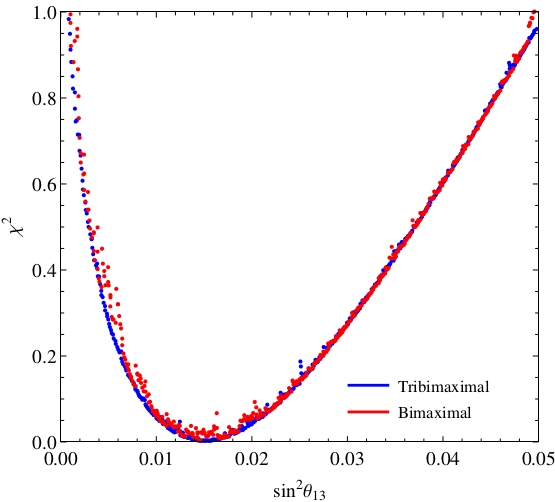}
\caption{\label{fig3}\it In the $SO(10)$ model the resulting $\chi^2$
  when starting in the TB or BM basis are equal within errors.  Note
  that the minimum $\chi^2$ value, $\chi^2 =0.003$, is obtained for
  $\sin^2\theta_{13} \sim 0.015$, just a bit below the measured value
  $\sin^2\theta_{13} \sim 0.022$.  Nevertheless, as the minimum
  $\chi^2$ is quite shallow for $\sin^2\theta_{13}<0.1$, the fit does not
  exhibit any strongly preferred value of $\theta_{13}$.}  \ec
\end{figure}

A closer look at the details of the fine tuning parameter reveals that
high $d_{FT}$ values are predominantly driven by the smallness of the
electron mass, combined with its extraordinary measurement
precision. Moreover, due to the presence of mixing, the $d_{FT}$
coming from, for instance, the 33 component of $h$, which is mainly
responsible for the top mass, is actually one of the largest
contributions to the global $d_{FT}$ (due to its contribution to the
electron mass) in both TB and BM scenarios. Although
this might be surprising at a first glance, we emphasize that the
dependence of the observables on the parameters is highly non trivial
due to the off-diagonal elements of the mass matrices.

In conclusion, as previously shown in Ref. \cite{blankenburg}, in this class of $SO(10)$ models one can obtain a reasonable fit to the data. Then one can reinterpret the result as BM corrected by the charged lepton diagonalization and 
explicitly determine the corrective terms arising from the fit. However the model does not imply BM mixing as a starting approximation. In fact, one could as well focus on 
TB mixing and make a similar interpretation. To predict, before diagonalization of charged leptons,  exact BM in the neutrino sector one would need additional dynamical ingredients. Independent of that, with the present value of  $\sin^2{\theta_{13}}$, a larger amount of fine tuning is needed in the BM case, as compared to the TB case, in order to reproduce 
the small values of the first generation masses. 

\begin{figure}[h!]
\bc
\includegraphics[scale=.5]{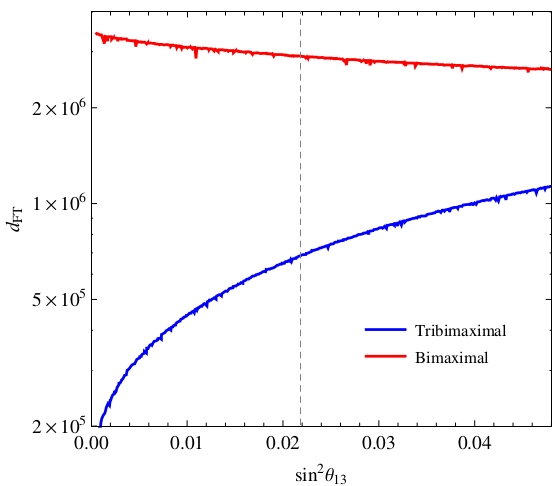}
\caption{\label{fig4}\it In the $SO(10)$ model the fine tuning parameter $d_{FT}$ increases (decreases) with $\sin^2{\theta_{13}}$ in the TB (BM) cases. For the physical value $\sin^2{\theta_{13}} \sim 0.022$ it is about 4 times larger in the BM case.}
\ec
\end{figure}

\section{Summary and Conclusion}
\label{sec:conc}

Models of neutrino mixing based on discrete flavour groups have been extensively studied. After the recent measurement of  $\sin^2{\theta_{13}}$
many of these models have been disfavoured, in particular among those aiming at implementing TB mixing. But models based on $S_4$ with BM mixing corrected by terms arising 
from the diagonalization of the charged lepton mass matrix remain as a viable and attractive possibility. In a GUT context, in these theories, it is also possible to implement 
the weak form of complementarity i.e. $\theta_{12}+\mathcal{O}(\theta_C)\sim \pi/4$ and to describe quark and lepton masses and mixings in a all comprehensive approach. 
Here we have discussed two examples of GUT models of BM, 
one based on $SU(5)$ and one on $SO(10)$. 
The $SU(5)$ model discussed here indeed has a broken flavour symmetry that contains $S_4$ and imposes 
the BM structure in the neutrino sector which is then corrected by terms arising from the diagonalization of charged lepton masses.
The $SO(10)$ model is based on Type-II see-saw and the origin of BM before diagonalization of charged leptons 
is in this case left unspecified. We have discussed the phenomenology of these models in the context of the most recent data and their relative merits. We have then compared these 
models based on a large symmetries with  models based on a minimum of symmetry where chance plays a central role, like Anarchy or models based on $U(1)_{FN}$. The $SU(5)$ 
model with  broken  $S_4$ symmetry emerges as the most viable and predictive theory.

\section*{Acknowledgments}
GA is very grateful to George Zoupanos and the Organising Committee
for inviting him at the Corfu Institute 2014 and for their kind
hospitality.  This research by GA was financed in part by the
LHCPHENONet and the Invisibles European Networks. The work of PM is
supported by an ESR contract of the European Union network FP7 ITN
INVISIBLES (Marie Curie Actions, PITN-GA-2011-289442).


\begin{thebibliography}{99}
 
\bibitem{Altarelli:2010gt} G.~Altarelli and F.~Feruglio, ,  Rev. Mod. Phys. {\bf 82} 2701  (2010) , arXiv:1002.0211.

\bibitem{ishikilu} H. Ishimori {\em et.~al.}, Prog. Theor. Phys. Suppl. 183, 1 (2010) , arXiv:1003.3552; 
S. F. King and C. Luhn, arXiv:1301.1340; S. F. King {\em et.~al.}, arXiv:1402.4271. 

\bibitem{grilu} W. Grimus and P. O. Ludl, J. Phys. A  45, 233001 (2012) , arXiv:1110.6376.

\bibitem{Hall:1999sn}
L.~J. Hall, H.~Murayama, and N.~Weiner,  Phys.
  Rev. Lett. {\bf 84}, 2572 (2000) ,  arXiv:hep-ph/9911341.

\bibitem{deGouvea:2003xe}
A.~de~Gouvea and H.~Murayama,  Phys. Lett.
  {\bf B573}, 94 (2003) , arXiv:hep-ph/0301050.

\bibitem{deGouvea:2012ac} A.~de~Gouvea and H.~Murayama,   arXiv:1204.1249.


\bibitem{Gonzalez-Garcia:2014bfa} 
  M.~C.~Gonzalez-Garcia, M.~Maltoni and T.~Schwetz,
  %``Updated fit to three neutrino mixing: status of leptonic CP violation,''
  JHEP {\bf 1411}, 052 (2014)
  [arXiv:1409.5439 [hep-ph]].


\bibitem{lessmin} E. Ma and D. Wegman, Phys. Rev. Lett. 107, 061803 (2011) , arXiv:1106.4269; S. F. King and C. Luhn, JHEP 09, 042 (2011) , 
arXiv:1107.5332;  F. Bazzocchi, arXiv:1108.2497; 
S.~Antusch, S.~F.~King, C.~Luhn and M.~Spinrath,
  %``Trimaximal mixing with predicted $\theta_{13}$ from a new type of constrained sequential dominance,''
  Nucl.\ Phys.\ B {\bf 856} (2012) 328
  [arXiv:1108.4278 [hep-ph]];
I.~de~Medeiros~Varzielas and L.~Merlo,   JHEP {\bf 02}, 062 (2011) , arXiv:1011.6662;
% W.~ Rodejohann and H.~ Zhang, arXiv:1207.1225; 
F.~Bazzocchi and L.~Merlo,
  %``Neutrino Mixings and the S4 Discrete Flavour Symmetry,''
  Fortsch.\ Phys.\  {\bf 61}, 571 (2013)
  [arXiv:1205.5135 [hep-ph]].

\bibitem{Lin:2009bw}
Y.~Lin,  Nucl. Phys. {\bf B824}, 95 (2010) , arXiv:0905.3534.

\bibitem{steve}  See, for example, 
% S.F. King, arXiv:1311.3295 
S.~Antusch and S.~F.~King,
  %``Sequential dominance,''
  New J.\ Phys.\  {\bf 6} (2004) 110
  [hep-ph/0405272].
\bibitem{altmix} R.~d.~A.~Toorop, F.~Feruglio and C.~Hagedorn,
Phys.\ Lett.\ B {\bf 703} (2011) 447 [arXiv:1107.3486 [hep-ph]]; Nucl.\ Phys.\ B{\bf 858}, 437 (2012) , arXiv:1112.1340; 
S.~F.~King, C.~Luhn and A.~J.~Stuart,
  %``A Grand Delta(96) x SU(5) Flavour Model,''
  Nucl.\ Phys.\ B {\bf 867} (2013) 203
  [arXiv:1207.5741 [hep-ph]]; 
C.~Hagedorn and D.~Meloni,
  %``D14 - A Common Origin of the Cabibbo Angle and the Lepton Mixing Angle theta^l_13,''
  Nucl.\ Phys.\ B {\bf 862} (2012) 691
  [arXiv:1204.0715 [hep-ph]];   
%   T. Araki et al, arXiv:1309.4217; 
 M.~Holthausen and K.~S.~Lim,
  %``Quark and Leptonic Mixing Patterns from the Breakdown of a Common Discrete Flavor Symmetry,''
  Phys.\ Rev.\ D {\bf 88} (2013) 033018
  [arXiv:1306.4356 [hep-ph]]; 
S.~F.~King, T.~Neder and A.~J.~Stuart,
  %``Lepton mixing predictions from $\Delta(6n^2)$ family Symmetry,''
  Phys.\ Lett.\ B {\bf 726} (2013) 312
  [arXiv:1305.3200 [hep-ph]];  
% G.-J. Ding and  S. F. King, arXiv:1403.5846.

\bibitem{lesssimm} S.~F.~Ge, D.~A.~Dicus and W.~W.~Repko,
Phys.\ Lett.\ B {\bf 702} (2011) 220
[arXiv:1104.0602 [hep-ph]]; 
Phys.\ Rev.\ Lett.\  {\bf 108} (2012) 041801
  [arXiv:1108.0964 [hep-ph]]; 
  D.~Hernandez and A.~Y.~Smirnov,
  %``Lepton mixing and discrete symmetries,''
  Phys.\ Rev.\ D {\bf 86} (2012) 053014
  [arXiv:1204.0445 [hep-ph]]; Phys.\ Rev.\ D {\bf 87} (2013) 5,  053005
  [arXiv:1212.2149 [hep-ph]].

\bibitem{cpfla} G.~Ecker, W.~Grimus and H.~Neufeld,
J.\ Phys.\ A {\bf 20} (1987) L807;  Int.\ J.\ Mod.\ Phys.\ A {\bf 3} (1988) 603;
W.~Grimus and M.~N.~Rebelo,
Phys.\ Rept.\  {\bf 281} (1997) 239
[hep-ph/9506272];
W.~Grimus and L.~Lavoura,
Phys.\ Lett.\ B {\bf 579} (2004) 113 [hep-ph/0305309];
% R. Krishnan, P. F. Harrison and W. G. Scott, arXiv:1211.2000; 
R.~N.~Mohapatra and C.~C.~Nishi,
Phys.\ Rev.\ D {\bf 86} (2012) 073007
[arXiv:1208.2875 [hep-ph]];
M.~Holthausen, M.~Lindner and M.~A.~Schmidt,
JHEP {\bf 1304} (2013) 122 [arXiv:1211.6953 [hep-ph]];
F.~Feruglio, C.~Hagedorn and R.~Ziegler,
JHEP {\bf 1307} (2013) 027
[arXiv:1211.5560 [hep-ph]]; Eur.\ Phys.\ J.\ C {\bf 74} (2014) 2753
[arXiv:1303.7178 [hep-ph]].

\bibitem{othercp} 
% G. C. Branco, J. M. Gerard and W. Grimus, Phys. Lett. B 136, 383 (1984) ; 
I.~de Medeiros Varzielas and D.~Emmanuel-Costa,
  %``Geometrical CP Violation,''
  Phys.\ Rev.\ D {\bf 84} (2011) 117901
  [arXiv:1106.5477 [hep-ph]];
% I. de Medeiros Varzielas, D. Emmanuel-Costa and P. Leser, Phys. Lett. B 716, 193 (2012) ; arXiv:1204.3633; 
% I. de Medeiros Varzielas, JHEP 1208, 055 (2012) ; arXiv:1205.3780; 
G.~Bhattacharyya, I.~de Medeiros Varzielas and P.~Leser,
Phys.\ Rev.\ Lett.\  {\bf 109} (2012) 241603
[arXiv:1210.0545 [hep-ph]]; 
% K. S. Babu and J. Kubo, Phys. Rev. D 71, 056006 (2005) ; arXiv:hep-ph/0411226]; 
% K. S. Babu, K. Kawashima and J. Kubo, Phys. Rev. D 83, 095008  (2011) , arXiv:1103.1664; 
M.~C.~Chen and K.~T.~Mahanthappa,
Phys.\ Lett.\ B {\bf 681} (2009) 444
[arXiv:0904.1721 [hep-ph]]; 
% A. Meroni, S. T. Petcov and M. Spinrath, Phys. Rev. D 86, 113003  (2012), arXiv:1205.5241;
G.~J.~Ding and S.~F.~King,
Phys.\ Rev.\ D {\bf 89} (2014) 9,  093020
[arXiv:1403.5846 [hep-ph]];
L.~L.~Everett, T.~Garon and A.~J.~Stuart,
arXiv:1501.04336 [hep-ph];
C.~C.~Li and G.~J.~Ding,arXiv:1503.03711 [hep-ph];
A.~Di Iura, C.~Hagedorn and D.~Meloni, arXiv:1503.04140 [hep-ph].

\bibitem{Ahn:2012nd}
  J.~K.~Ahn {\it et al.}  [RENO Collaboration],
  %``Observation of Reactor Electron Antineutrino Disappearance in the RENO Experiment,''
  Phys.\ Rev.\ Lett.\  {\bf 108} (2012) 191802
  [arXiv:1204.0626 [hep-ex]].
  
\bibitem{Abe:2012tg} 
  Y.~Abe {\it et al.}  [Double Chooz Collaboration],
  %``Reactor electron antineutrino disappearance in the Double Chooz experiment,''
  Phys.\ Rev.\ D {\bf 86}, 052008 (2012)
  [arXiv:1207.6632 [hep-ex]].
  
\bibitem{An:2013zwz} 
  F.~P.~An {\it et al.}  [Daya Bay Collaboration],
  %``Spectral measurement of electron antineutrino oscillation amplitude and frequency at Daya Bay,''
  Phys.\ Rev.\ Lett.\  {\bf 112}, 061801 (2014)
  [arXiv:1310.6732 [hep-ex]].

\bibitem{Abe:2013hdq} 
  K.~Abe {\it et al.}  [T2K Collaboration],
  %``Observation of Electron Neutrino Appearance in a Muon Neutrino Beam,''
  Phys.\ Rev.\ Lett.\  {\bf 112}, 061802 (2014)
  [arXiv:1311.4750 [hep-ex]].  

\bibitem{Barger:1998ta} 
  V.~D.~Barger, S.~Pakvasa, T.~J.~Weiler and K.~Whisnant,
  %``Bimaximal mixing of three neutrinos,''
  Phys.\ Lett.\ B {\bf 437}, 107 (1998)
  [hep-ph/9806387]. 
 
 \bibitem{Mohapatra:1998ka} 
  R.~N.~Mohapatra and S.~Nussinov,
  %``Bimaximal neutrino mixing and neutrino mass matrix,''
  Phys.\ Rev.\ D {\bf 60}, 013002 (1999)
  [hep-ph/9809415].

\bibitem{Altarelli:2009gn} 
  G.~Altarelli, F.~Feruglio and L.~Merlo,
  %``Revisiting Bimaximal Neutrino Mixing in a Model with S(4) Discrete Symmetry,''
  JHEP {\bf 0905}, 020 (2009)
  [arXiv:0903.1940 [hep-ph]].
  
  
\bibitem{Meloni:2011fx}
  D.~Meloni,
  %``Bimaximal mixing and large theta13 in a SUSY SU(5) model based on S4,''
  JHEP {\bf 1110} (2011) 010
  [arXiv:1107.0221 [hep-ph]].


 
\bibitem{Raidal:2004iw} 
  M.~Raidal,
  %``Relation between the neutrino and quark mixing angles and grand unification,''
  Phys.\ Rev.\ Lett.\  {\bf 93}, 161801 (2004)
  [hep-ph/0404046].
  
  
\bibitem{Minakata:2004xt} 
  H.~Minakata and A.~Y.~Smirnov,
  %``Neutrino mixing and quark-lepton complementarity,''
  Phys.\ Rev.\ D {\bf 70}, 073009 (2004)
  [hep-ph/0405088].
  
\bibitem{Frampton:2004vw} 
  P.~H.~Frampton and R.~N.~Mohapatra,
  %``Possible gauge theoretic origin for quark-lepton complementarity,''
  JHEP {\bf 0501}, 025 (2005)
  [hep-ph/0407139].
  
  
\bibitem{Antusch:2005ca} 
  S.~Antusch, S.~F.~King and R.~N.~Mohapatra,
  %``Quark-lepton complementarity in unified theories,''
  Phys.\ Lett.\ B {\bf 618}, 150 (2005)
  [hep-ph/0504007].
  
\bibitem{Patel:2010hr} 
  K.~M.~Patel,
  %``An SO(10)XS4 Model of Quark-Lepton Complementarity,''
  Phys.\ Lett.\ B {\bf 695}, 225 (2011)
  [arXiv:1008.5061 [hep-ph]].
  
  
\bibitem{blankenburg}
G.~Altarelli and G.~Blankenburg,
  %``Different $SO(10)$ Paths to Fermion Masses and Mixings,''
  JHEP {\bf 1103} (2011) 133
  [arXiv:1012.2697 [hep-ph]].

\bibitem{Froggatt:1978nt}
  C.~D.~Froggatt and H.~B.~Nielsen,
  %``Hierarchy of Quark Masses, Cabibbo Angles and CP Violation,''
  Nucl.\ Phys.\ B {\bf 147} (1979) 277.

\bibitem{AFMM:2012}
G.~Altarelli, F.~Feruglio, I.~Masina and L.~Merlo,
  %``Repressing Anarchy in Neutrino Mass Textures,''
  JHEP {\bf 1211} (2012) 139
  [arXiv:1207.0587 [hep-ph]].

\bibitem{melmer} J.~Bergstrom, D.~Meloni and L.~Merlo,
  %``Bayesian comparison of U(1) lepton flavor models,''
  Phys.\ Rev.\ D {\bf 89} (2014) 9,  093021
  [arXiv:1403.4528 [hep-ph]].


\bibitem{5DSU5}
E.~Witten,
  %``Symmetry Breaking Patterns In Superstring Models,''
  Nucl.\ Phys.\  B {\bf 258} (1985) 75;
  %%CITATION = NUPHA,B258,75;%%
Y.~Kawamura,
  %``Triplet-doublet splitting, proton stability and extra dimension,''
  Prog.\ Theor.\ Phys.\  {\bf 105} (2001) 999
  [arXiv:hep-ph/0012125];
  %%CITATION = PTPKA,105,999;%%
A.~E.~Faraggi,
  %``Doublet-triplet splitting in realistic heterotic string derived models,''
  Phys.\ Lett.\  B {\bf 520} (2001) 337
  [arXiv:hep-ph/0107094] and references therein.
  %%CITATION = PHLTA,B520,337;%%

\bibitem{5D}
L.~J.~Hall and Y.~Nomura,
  %``Gauge unification in higher dimensions,''
  Phys.\ Rev.\  D {\bf 64} (2001) 055003
  [arXiv:hep-ph/0103125];
  %%CITATION = PHRVA,D64,055003;%%
Y.~Nomura,
  %``Strongly coupled grand unification in higher dimensions,''
  Phys.\ Rev.\  D {\bf 65} (2002) 085036
  [arXiv:hep-ph/0108170];
  %%CITATION = PHRVA,D65,085036;%%
L.~J.~Hall and Y.~Nomura,
  %``A complete theory of grand unification in five dimensions,''
  Phys.\ Rev.\  D {\bf 66} (2002) 075004
  [arXiv:hep-ph/0205067].
  %%CITATION = PHRVA,D66,075004;%%
  
\bibitem{5Dfreedom}
G.~Altarelli and F.~Feruglio,
  %``SU(5) grand unification in extra dimensions and proton decay,''
  Phys.\ Lett.\  B {\bf 511} (2001) 257
  [arXiv:hep-ph/0102301];
  %%CITATION = PHLTA,B511,257;%%
A.~Hebecker and J.~March-Russell,
  %``A minimal S(1)/(Z(2) x Z'(2)) orbifold GUT,''
  Nucl.\ Phys.\  B {\bf 613} (2001) 3
  [arXiv:hep-ph/0106166];
  %%CITATION = NUPHA,B613,3;%%
A.~Hebecker and J.~March-Russell,
  %``The flavour hierarchy and see-saw neutrinos from bulk masses in 5d
  %orbifold GUTs,''
  Phys.\ Lett.\  B {\bf 541} (2002) 338
  [arXiv:hep-ph/0205143].
  %%CITATION = PHLTA,B541,338;%%

\bibitem{Altarelli:2008bg}
  G.~Altarelli, F.~Feruglio, C.~Hagedorn,
  %``A SUSY SU(5) Grand Unified Model of Tri-Bimaximal Mixing from A(4),''
  JHEP {\bf 0803}, 052-052 (2008).
  [arXiv:0802.0090 [hep-ph]].  
  
\bibitem{Joshipura:2011nn} 
  A.~S.~Joshipura and K.~M.~Patel,
  %``Fermion Masses in SO(10) Models,''
  Phys.\ Rev.\ D {\bf 83}, 095002 (2011)
  [arXiv:1102.5148 [hep-ph]].


  
 \bibitem{Altarelli:2002sg}
  G.~Altarelli, F.~Feruglio and I.~Masina,
  %``Models of neutrino masses: Anarchy versus hierarchy,''
  JHEP {\bf 0301} (2003) 035
  [hep-ph/0210342].  
  
  
\bibitem{Altarelli:2012ia}
  G.~Altarelli, F.~Feruglio, I.~Masina and L.~Merlo,
  %``Repressing Anarchy in Neutrino Mass Textures,''
  JHEP {\bf 1211} (2012) 139
  [arXiv:1207.0587 [hep-ph]].
  
\bibitem{tII}
  G.~Lazarides, Q.~Shafi and C.~Wetterich,
  Nucl.\ Phys.\  B {\bf 181}, 287 (1981);
  J.~Schechter and J.~W.~F.~Valle,
  Phys.\ Rev.\  D {\bf 22}, 2227 (1980);
  R.~N.~Mohapatra and G.~Senjanovic,
  Phys.\ Rev.\  D {\bf 23}, 165 (1981).



\bibitem{tI}
P.~Minkowski, Phys.\ Lett.\ B {\bf 67} (1977) 421;
T.~Yanagida in {\em Workshop on Unified Theories, KEK Report 79-18},
p.~95, 1979;
M.~Gell-Mann, P.~Ramond and R.~Slansky, {\em Supergravity}, p.~315,
Amsterdam: North Holland, 1979;
S.~L. Glashow, {\em 1979 Cargese Summer Institute on Quarks and
Leptons}, p.~687,
New York: Plenum, 1980;
R.~N. Mohapatra and G.~Senjanovic,
Phys.\ Rev.\ Lett,\ {\bf 44} (1980) 912.

\bibitem{mimura}
  B.~Dutta, Y.~Mimura and R.~N.~Mohapatra,
  Phys.\ Rev.\ Lett.\  {\bf 94}, 091804 (2005)
  [hep-ph/0412105],
  Phys.\ Rev.\  D {\bf 72}, 075009 (2005)
  [hep-ph/0507319],
  Phys. \ Rev. {\bf D80} (2009) 095021 
  [ArXiv:0910.1043 [hep-ph]].

  
  
\bibitem{goh}
H.~S.~Goh, R.~N.~Mohapatra and S.~P.~Ng,
        Phys.\ Lett.\  B {\bf 570}, 215 (2003)
        [hep-ph/0303055],
        Phys.\ Rev.\  D {\bf 68}, 115008 (2003)
        [hep-ph/0308197];
H.~S.~Goh, R.~N.~Mohapatra and S.~Nasri,
        Phys.\ Rev.\ {\bf D70} 075022 (2004)
        [hep-ph/0408139].

\bibitem{aulak1}
C.~S.~Aulakh, B.~Bajc, A.~Melfo, G.~Senjanovic and F.~Vissani,
        Phys.\ Lett.\  B {\bf 588}, 196 (2004)
        [hep-ph/0306242].

\bibitem{aulak2}
C.~S.~Aulakh and S.K~ Garg, 
        Nucl.\ Phys.\  {\bf B757} 47 (2006) 
        [hep-ph/0512224], 
        [hep-ph/0612021].
        
        
\bibitem{bajc1}
B. Bajc, A. Melfo, G. Senjanovic, F. Vissani,
        Phys. \ Lett. \ {\bf B634} (2006)272
        [ArXiv:0511352 [hep-ph]].
        
        
\bibitem{melfo}
A.~Melfo, A.~Ramirez and G.~Senjanovic,
Phys.\ Rev.\ {\bf D82} (2010) 075014
[ArXiv:1005.0834 [hep-ph]].        
  
  
\end{thebibliography}
\end{document}